      [1999/12/01 v1.4c Il Nuovo Cimento]
\documentclass{cimento}

\usepackage{graphicx}
\usepackage{bm}

\title{TMDs, universality and factorization}
\author{U.~D'Alesio\from{ins:ca}\thanks{Talk delivered at the ``Third International Workshop on Transverse Polarization Phenomena in Hard Scattering'' (Transversity 2011), 29 Aug.-2 Sep.~2011, Veli Lo\u{s}inj (Croatia).}}
\instlist{\inst{ins:ca} Dipartimento di Fisica, Universit\`a di Cagliari, and INFN, Sezione di Cagliari, Cittadella Universitaria, Monserrato (CA), Italy
}
\PACSes{\PACSit{12.38.-t}{Quantum Chromodynamics}
\PACSit{13.88.+e}{Polarization in interactions and scattering}
\PACSit{12.39.St}{Factorization}}
\begin{document}

\maketitle

\vspace*{-0.7cm}

\begin{abstract}
We present a short overview on transverse momentum dependent parton distribution and fragmentation functions, giving their partonic interpretation and ways to access them. We then discuss the issue of their universality and its connection to factorization in perturbative QCD.
\end{abstract}

\vspace*{-.75cm}

\section*{Introduction}
The description of the internal structure of nucleons in terms of their fundamental constituents, quarks and gluons, has been a challenging issue since the formulation of Quantum Chromodynamics (QCD).
Deep inelastic scattering (DIS) experiments of high energy leptons off nucleons have played a central role in this respect, allowing to reach a very accurate comprehension of the nucleon structure and, at the same time, test QCD in the perturbative regime.
The information so obtained is encoded in the parton distribution function (PDF) $f_1^a(x)$, i.e.~the number density of partons of type $a$ carrying a momentum fraction $x$ of the nucleon momentum. Analogously we have reached a good comprehension of the number density of longitudinally polarized partons inside longitudinally polarized nucleons, encoded in the helicity distribution $g_1^a(x)$. Both these distributions, as predicted by QCD, depend on the large scale involved in the reaction.

Although well established, the understanding of the nucleon structure so obtained is basically one-dimensional and, therefore, not complete. In such a picture many aspects remain to be addressed: the spatial distribution of partons inside a nucleon, their motion in the transverse plane, the parton contribution to the nucleon orbital angular momentum and the potential correlations between orbital motion and quark or nucleon spins.
In the last two decades these issues have received a lot of attention, theoretical progress has put many of them on a solid field-theoretical basis and, thanks to significant experimental achievements, a new and more complete picture of the nucleon structure is emerging.

Two complementary descriptions can be combined to provide tomography images of the inner structure of the nucleon: the description of partons in the transverse plane in momentum space, encoded in the transverse momentum dependent parton distributions (TMDs); the description of partons in the transverse plane in coordinate space, encoded in the generalized parton distributions (GPDs)~\cite{GPD}. Here we will focus on the first topic.

The study of TMDs involves very deep issues: most TMDs would vanish in absence of parton orbital angular momentum and are due to couplings of the parton transverse momentum to the nucleon/quark spin. Even if there are other sources of information, like the measurement of nucleons' anomalous magnetic moments, TMDs would eventually allow us to map the parton orbital angular momentum flavor by flavor. 
Moreover, the existence of some TMDs is related to fundamental properties of QCD, mainly its color gauge invariance, and calculable process dependencies leading to a universality breaking are expected. This would represent a crucial test for our understanding of QCD at work.

\section{What are TMDs and where and how can we learn on them?}
\label{TMDsgeneral}
The simplest TMD, in the parton model, is the unintegrated unpolarized parton distribution $f_1^a(x,\bm{k}_\perp)$, giving the number density of partons with a fraction $x$ of the nucleon momentum and transverse momentum $\bm{k}_\perp$. The motivation behind this TMD traces back to the parton model analysis of the Drell-Yan (DY) process, $h_A h_B\to l^+l^- X$, where the (low) transverse momentum, $\bm{q}_T$, of the final lepton pair originates from the transverse momenta of the two initial partons (a $q\bar q$ pair):
if the partons had no transverse momentum, the cross section would be a delta-function at $\bm{q}_T=0$, against the experimental observation. In this sense TMDs can be seen as a generalization of the collinear PDFs.

In a more formal way, TMDs are defined via the unintegrated quark-quark correlator for a
polarized nucleon (with spin $\bm{S}$) entering the DIS process
\begin{equation} \label{eq:corr}
  \Phi_{ij}^q(x,\mbox{\boldmath $k$}_\perp,\mbox{\boldmath $S$})_\eta
  = \int \frac{dz^-d^2z_\perp}{(2 \pi)^3}\,
  \mathrm{e}^{ik\cdot z}\langle \mbox{\boldmath $P$},\mbox{\boldmath $S$}\,|
  \,\bar{\psi}_j^q(0) \, \mathcal{W}_\eta(0,z) \, \psi_i^q(z)\,
  |\mbox{\boldmath $P$},\mbox{\boldmath $S$}\rangle \Big|_{z^+=0}\>,
\end{equation}
where the gauge link operator $\mathcal{W}_\eta(0,z)$ ensures the color gauge invariance of the matrix element and can depend on the process (via the path $\eta$), as further discussed.
By proper Dirac projections one can extract the eight leading-twist TMDs for the  nucleon~\cite{Kotzinian:1994dv,Mulders:1995dh,Boer:1997nt}:
\begin{eqnarray}
  \frac{1}{2}\,\mathrm{tr}\left[
  \gamma^{+}\,\Phi^q(x,\mbox{\boldmath $k$}_\perp,
  \mbox{\boldmath $S$}) \,\right] \label{f:funct}
  &=& f_{1}^q(x,k_\perp)-\frac{\varepsilon ^{jk}\,k_\perp^{j}\,S_T^{k}}{M}
      \,f_{1T}^{\perp q}(x,k_\perp) \\
   \frac{1}{2}\,\mathrm{tr}\left[
  \gamma^{+}\gamma_{5}\,\Phi^q(x,\mbox{\boldmath $k$}_\perp,
  \mbox{\boldmath $S$})\,\right]\label{g:funct}
  &=& S_{L}\,g_{1}^q(x,k_\perp)+\frac{\mbox{\boldmath $k$}_\perp\cdot\mbox{\boldmath $S$}_{T}}{M}
      g_{1T}^q(x,k_\perp) \\
  \hspace{-8mm}
  \frac{1}{2}\,\mathrm{tr}\left[ i\sigma ^{j+}\gamma _{5} \,
  \Phi^q (x,\mbox{\boldmath $k$}_\perp,\mbox{\boldmath $S$})\right]
  &=& S_{T}^{j}\,h_1^q(x,k_\perp)
   +S_{L}\,\frac{k_\perp^{j}}{M}\,h_{1L}^{\perp q}(x,k_\perp) \nonumber\\
   &+&\label{h:funct}
   \frac{(k_\perp^{j}\,k_\perp^{k}-\frac{1}{2}\,\mbox{\boldmath $k$}_\perp^{\:2}\,
     \delta^{jk})S_{T}^{k}}{M^{2}}\,h_{1T}^{\perp q}(x,k_\perp)
   +\frac{\varepsilon^{jk}\,k_\perp^{k}}{M}\,h_1^{\perp q}(x,k_\perp).
\end{eqnarray}
The gamma-structures signal the quark polarizations: $\gamma^+$ for unpolarized quarks, $\gamma^+\gamma_5$ singles out longitudinally polarized quarks, and finally $i\,\sigma^{j+}\gamma_5$ selects transversely polarized quarks.
Notice that eight are exactly the parity-conserving combinations that can be built from the two spin pseudo-vectors ($\bm{s}_q$ for the quark, $\bm{S}$ for the nucleon) and the two vectors $\bm{k}_\perp$ (the quark transverse momentum) and $\bm{P}$ (the nucleon momentum).

Let us now discuss their partonic interpretation, starting with the three TMDs that survive in the collinear limit: $f_1^q(x, k_\perp)$ is the unpolarized, $k_\perp$-dependent quark distribution; $g_{1}^q(x, k_\perp)$ is the $k_\perp$-dependent helicity distribution; $h_1^q(x, k_\perp)$ is the analogue of the helicity distribution, for transverse nucleon spin, {\it i.e.} the transversity distribution. This last one being chiral-odd, decouples from DIS and turns out harder to measure. So far only one extraction of the $u$ and $d$ quark transversities is available in the literature~\cite{Anselmino:2007fs}, obtained by a combined fit of semi-inclusive deep inelastic scattering (SIDIS) and $e^+e^-$ data.
The other TMDs share a very special feature: they would vanish in the absence of orbital angular momentum due to angular momentum conservation.

$f_{1T}^{\perp q}$, in Eq.~(\ref{f:funct}), is the Sivers function~\cite{Sivers:1989cc}, also denoted as $ \Delta^N\! f_{q/p^\uparrow}(x,k_\perp)$ (see Refs.~\cite{Anselmino:2005sh,Bacchetta:2004jz}), giving the asymmetric part of the distribution of unpolarized partons inside a (transversely) polarized proton:
\begin{equation}
f_{1}^q(x, \bm{k}_\perp;\bm{0},\bm{S}) =
f_1^{q}(x, k_\perp) - \frac{k_\perp}{M} \,  f_{1T}^{\perp q}(x,k_\perp) \> \bm{S} \cdot  (\hat{\bm{P}} \times \hat{\bm{k}}_\perp) \> .
\end{equation}

$h_{1}^{\perp q}$, in Eq.~(\ref{h:funct}) is the Boer-Mulders function~\cite{Boer:1997nt}, giving the asymmetric part of the distribution of (transversely) polarized quarks inside an unpolarized proton:
\begin{equation}
f_1^{q}(x, \bm{k}_\perp;\bm{s}_q, \bm{0})
= \frac 12 \,\Big[ f_1^{q}(x, k_\perp) - \frac{k_\perp}{M}
\, h_{1}^{\perp q}(x, k_\perp) \> \bm{s}_q \cdot
(\hat{\bm{P}} \times \hat{\bm{k}}_\perp)\Big] \;.
\end{equation}
The origin and expected process dependence of these two TMDs are related to fundamental QCD effects, as we will discuss in the next section.

$h_{1T}^{\perp q}$, in Eq.~(\ref{h:funct}), gives the transverse polarization of quarks orthogonal to the nucleon transverse polarization;
$g_{1T}^{q}$, $h_{1L}^{\perp q}$, in Eqs.~(\ref{g:funct}), (\ref{h:funct}), also named {\em worm-gear} functions, represent the amount of longitudinally/transversely polarized quarks in a transversely/longitudinally polarized proton. Note that they are the real parts of interference amplitudes whose imaginary parts are, respectively, $f_{1T}^\perp$ and $h_1^\perp$.

An analogous pattern can be defined for gluon TMDs~\cite{Mulders:2000sh,Anselmino:2005sh}, where instead of transverse spin we will have to consider linear polarizations.

In the fragmentation sector, for a spin-1/2 hadron there are eight TMD quark/gluon fragmentation functions (FFs) following the same scheme as that for TMD PDFs. For a spinless (or unpolarized) hadron only two leading-twist TMD FFs exist: the unpolarized $k_\perp$-dependent parton FF and the Collins function~\cite{Collins:1992kk}, $H_1^{\perp q}$, giving the asymmetric part of the distribution of unpolarized hadrons in the fragmentation of a (transversely) polarized quark. This is chiral odd and represents the ideal partner of transversity.

\begin{figure}
\center
\includegraphics[width=8cm]{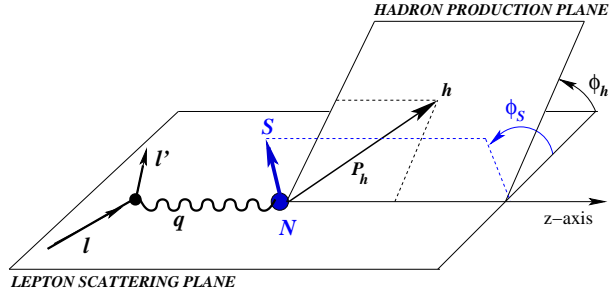}
\caption{Kinematics for SIDIS. $\phi_S$ ($\phi_h$) is the azimuthal angle of the transverse spin (final hadron transverse momentum) w.r.t.~the lepton scattering plane. }
\label{sidis}
\end{figure}

Concerning the ways one can learn on TMDs, SIDIS process, $lN\to l'hX$, where, at variance with DIS, one also observes a hadron in the final state (see Fig.~\ref{sidis}), is definitely the main source of information. By choosing the polarizations of the incoming lepton and the target nucleon and by looking at different azimuthal dependencies~\cite{Kotzinian:1994dv,Bacchetta:2006tn,Anselmino:2011ch}, one can have a complete access to TMDs in terms of simple factorized expressions. This is strictly true in a parton model picture (tree-level approximation), where soft gluon exchanges are neglected (see next section) and in the region of low transverse momentum of the final hadron ($P_{hT}\simeq \Lambda_{\rm QCD}$). Notice that all existing extractions of TMDs are based on this approximation.
For its relevance we mention here the case of SIDIS with unpolarized leptons on transversely polarized nucleon.
In this case the asymmetry exhibits a $\sin(\phi_h-\phi_S)$, a $\sin(\phi_h+\phi_S)$ and a $\sin(3\phi_h-\phi_S)$ modulation involving, respectively, the Sivers function, the Collins function convoluted with the transversity distribution and the Collins function convoluted with $h_{1T}^{\perp q}$. In particular, the observation at HERMES~\cite{hermes:exp} of the first two modulations led to the unambiguous evidence of the Sivers and the Collins effects. Present data come from three polarized fixed-target experiments: HERMES (DESY), COMPASS (CERN)~\cite{compass:exp} and HallA (JLab)~\cite{Gao:2010av}. The planned EIC project, a high-energy, high-luminosity, polarized electron-proton collider would undoubtedly be a precision machine for the study of TMDs~\cite{Anselmino:2011ay}.

Complementary and crucial information on TMDs, in view of global fits (see, e.g., Ref.~\cite{Melis:2011}), can be obtained from DY processes and $e^+e^-$ annihilations into an almost back-to-back hadron pair. In particular, the observation of single spin asymmetries in DY scattering, still missing, would play a fundamental role in our comprehension of TMDs in QCD. The ongoing and future programs at J-PARC, RHIC, COMPASS and PAX could definitely help in this respect.
Belle experiment~\cite{belle:exp} have provided very accurate data sets on the Collins effect in $e^+e^-$ processes and the BaBar collaboration is currently working on a similar analysis~\cite{Garzia:2011}. See Ref.~\cite{Barone:2010zz} for an up-to-date review.

Gluon TMDs have recently received special attention. In particular, it has been shown how the distribution of linearly polarized gluons inside an unpolarized proton, $h_1^{\perp g}$, can be directly accessed in $ep\to e' Q\bar Q X$~\cite{Boer:2010zf} and in $pp\to\gamma\gamma X$~\cite{Qiu:2011ai}. See also Sect.~\ref{smallx}.

\section{Universality and factorization}
\label{univ-fact}
QCD is a local gauge invariant field theory and consistently a gauge invariant definition of TMDs is mandatory. This can be obtained by the insertion in Eq.~(\ref{eq:corr}) of a gauge link (Wilson line) -- a path ordered exponential of gauge field -- filling the space separation between the quark fields. Even if true also for ordinary collinear PDFs, there are two crucial aspects that characterize the role of a Wilson line in TMDs: $i)$ some TMDs are non-zero only if the Wilson line is included; $ii)$ the Wilson line can depend on the process leading to a nontrivial universality behaviour of TMDs.

Let us start with the standard collinear PDF, that, in the parton model, can be interpreted as the hadron expectation value of the number density of partons. When we move to QCD, this interpretation is still valid if one chooses the light-cone gauge ($A^+=0$, with the Wilson line becoming unity). On the other hand this choice leads to a singularity in the gluon propagator that must be properly regularized in order to carry out the factorization. Alternatively, one can use a covariant gauge (e.g.~the Feynman gauge): this avoids the singularity problem mentioned above, at the price of having to consider extra regions (not anymore sub-leading), involving collinear gluons.

For a collinear PDF, where the quark fields are separated along a light-like minus direction, the Wilson line follows a {\em straight} path. For a TMD, where parton fields are no longer at light-like separation, the Wilson line has to make a detour at infinity in the {\em transverse} direction. Notice that a Wilson line is a resummation of extra gluons in the derivation of factorization and therefore its structure is controlled by this requirement, that is by the procedure of separating the hard from the soft part. For a systematic study of the role of the gauge links in TMDs see also Refs.~\cite{Ji:gaugelinks,Ji:2004wu}.

TMD factorization has been proved for the following processes: $l N \to l^\prime  h  X$ (SIDIS), $l^+ l^- \to h_1  h_2  X$ (annihilation process) and $h_1  h_2 \to l^+ l^-  X$ (DY). In all these cases there is, at leading order, one hard subprocess (a virtual line coupling to a parton line): this implies a \emph{simple} color flow from the hard to the soft part. Which gauge-link is relevant in a particular non-perturbative function depends on the color flow in the full process. In particular, if the color flows after the hard scattering, as in SIDIS, the gauge link is future pointing, if the color is annihilated before the hard scattering, as in DY, the gauge link is past pointing.
Parity and time-reversal invariance allow to relate these different paths: what turns out is that six TMDs (those even under time-reversal) are universal, while two, naively time-reversal odd (T-odd), are not universal. However, this non-universality is under control and calculable, consisting simply in a sign change~\cite{Collins:2002kn}:
\begin{equation} \label{eq:univ}
f_{1T}^\perp \big|_{\rm{DY}} = - f_{1T}^\perp \big|_{\rm{DIS}} \,,
\qquad \qquad h_{1}^\perp \big|_{\rm{DY}} = - h_{1}^\perp
\big|_{\rm{DIS}} \,.
\end{equation}
The observation of the Sivers effect in DY is therefore crucial and a test of the predicted sign change an outstanding issue in hadronic physics.

For the case of TMD FFs the shape of the Wilson line appears to have no influence on physical observables, leading to a standard universality behaviour~\cite{Collins:2004nx}.

In the TMD factorization procedure, however, extra complications arise.
In fact, gluons moving with negative infinity rapidity w.r.t.~the parent nucleon lead to (rapidity) divergences that cannot be canceled between real and virtual gluon contributions (as happens for collinear PDFs)~\cite{Collins:2003.08}. One option to regularize them is to tilt the gauge link out of the light-like direction. This implies the introduction of a new parameter with a consequent generalization of the renormalization procedure, leading to the Collins Soper Sterman (CSS) evolution equations~\cite{CS,Collins:1984kg}.
Another complication comes from Wilson-line self-energies, present in the naive definition of TMDs, but not in the actual cross section: these create further divergences~\cite{Bacchetta:2008xw}.
Finally, a complete treatment of TMD factorization, beyond the tree-level approximation, involves soft gluons, which give rise to an extra soft factor in the factorization formula~\cite{Collins:1984kg,Ji:2004wu}.

Among the alternative definitions of TMDs, based on different ways to handle soft and rapidity divergences in the derivation of factorization, we recall here two of them.
Adopting the Feynman gauge, Collins has recently shown~\cite{Collins:2011zzd} that the soft factor, that resums soft gluons in the TMD factorization formula of SIDIS (DY), can be properly included in the re-definition of TMDs. This avoids the explicit presence of a non-perturbative quantity that cannot be determined independently from data, somehow spoiling the simple picture of having soft parts only in PDFs and/or FFs.
Rapidity and Wilson-line self-energy divergences cancel leading to a TMD factorization with maximal universality for the TMDs; the evolution equations, in the rapidity cut-off, turn out homogeneous. See Ref.~\cite{Aybat:2011zv} for a first phenomenological application of this prescription.

In the light-like axial gauge, Cherednikov and Stefanis~\cite{Igor} proposed an alternative definition of TMDs, at one-loop order approximation with a regularized gluon propagator. In such a gauge, the transverse gauge link at infinity (absent in covariant gauges) is responsible for T-odd TMDs. Evolution equations, following a CSS-like procedure, are provided even if a complete proof of factorization is still missing.

At present it is not clear if these different definitions are all formally acceptable.

\subsection{TMD factorization breaking effects}
\label{fact-break}
The discussion above refers to processes where TMD factorization is known to hold. There are also, however, situations, where TMD-factorization is expected to fail~\cite{Piet,Collins:2007nk,Bacchetta:2007wc,Rogers:2010dm}. The key issue is the failure of the usual Ward identity arguments that ordinarily allow to factorize gluons into Wilson lines.
In processes like hadron production in hadron-hadron collisions, $h_Ah_B\to h_C h_D X$, where the final particles are hadrons or jets, the color flow has a more complicated structure due to the presence of many colored partons. Therefore, the simple redefinition of TMDs working for SIDIS and DY processes is not possible anymore and a violation of universality is expected. A detailed analysis~\cite{Rogers:2010dm} of multiple gluons in the derivation of TMD factorization shows that even a sort of ``generalized'' TMD factorization scheme (a factorized cross section into a hard part and well-defined, albeit non-universal, matrix elements for each hadron) seems not possible. Therefore, the problem with factorization in the hadron-production of hadrons seems more than just a problem with universality.

\subsection{TMDs at small $x$}
\label{smallx}
Recent works~\cite{Xiao,Dominguez} have extended the discussion on the universality and factorization issues to the small-$x$ domain, where the $k_\perp$-dependent distributions have been commonly used to describe the relevant physics phenomena. In particular, it is well know that the $k_\perp$-dependent unpolarized gluon distribution plays a central role in small $x$ saturation phenomena. Due to the presence of a semi-hard scale (the saturation scale), an effective theory, the color glass condensate (CGC), was proposed to systematically study this physics (see Refs.~\cite{CGC} and references therein). Two $k_\perp$-dependent unpolarized gluon distributions are widely used: $i)$ the Weizs\"acker-Williams (WW) distribution~\cite{McLerran}, describing the gluon number density in CGC formalism; $ii)$ the so-called color-dipole distribution (more precisely, its Fourier transform), entering, e.g., the calculations of inclusive particle production in $pA$ collisions~\cite{CGC}.

In Refs.~\cite{Dominguez} the authors establish an effective TMD factorization at small $x$ in the correlation limit, where the transverse momentum imbalance of, e.g., two outgoing jets, produced in the hard scattering of a dilute probe off dense nuclei, is much smaller than the individual transverse jet momenta.
After proving the equivalence of the TMD and CGC approaches in their overlapping domain, they show how the WW gluon distribution is directly probed in the quark-antiquark jet correlation in DIS, whilst the color-dipole distribution is probed in the direct photon-jet correlation in $pA$ collisions.
Metz and Zhou extended this analysis to the case of $h_1^{\perp g}$ for a large nucleus, showing that the WW distribution, probed in $\gamma^*A \to {\rm jet}\, {\rm jet}\,X$, saturates its positivity bound at high $k_\perp$, while being suppressed at low $k_\perp$; in contrast, the dipole distribution, probed in $pA\to \gamma^*\, {\rm jet}\, X$, saturates the bound for any value of $k_\perp$~\cite{Metz:2011wb}.

\subsection{More on TMDs in hadron-hadron processes}
\label{TMDs-pp}
We consider here the case of single spin asymmetries (SSAs) for single inclusive hadron production in hadron-hadron collisions. Generally defined as the ratio of the difference and the sum of the cross sections when the hadron's spin vector $S_T$ is flipped, the SSAs in $pp\to \pi X$ are among the earliest processes studied~\cite{Bunce:1976yb} and remain extremely challenging to explain.
The trend of large SSAs in the pioneering fixed target experiments has been observed over a wide range of energies up to those reachable at Relativistic Heavy Ion Collider (RHIC)~\cite{STAR:exp}.

Two approaches have been proposed in the framework of perturbative QCD to account for these effects: $i)$ a collinear factorization formalism at next-to-leading-power (twist-three) in the hard scale, where SSAs are given by a convolution of universal non-perturbative quark-gluon-quark correlation functions and hard scattering amplitudes~\cite{twist3:theo}; $ii)$ a phenomenological approach based on factorization in terms of a hard scattering cross section and TMDs, the so-called generalized parton model (GPM)~\cite{GPM:theo,Anselmino:2005sh}. See also Ref.~\cite{D'Alesio:2007jt} for a review on this.
Here factorization is assumed as a reasonable starting point together with universal leading-twist naive time-reversal odd TMD PDFs. A satisfactory description of SSA data has been shown to be possible~\cite{Boglione:2007dm}.

It is worth to mention that there is a quantitative connection between the first moment of the Sivers function and the twist-three correlation function~\cite{Boer:2003cm}. However, a reanalysis of SSAs in the twist-three approach~\cite{Kang:2011hk} has raised a new puzzling result: a sign mismatch between phenomenological extractions in SIDIS and $pp\to\pi X$ (see~\cite{Vogelsang:2011} for details).

Recently, an extension of the GPM including color gauge links has been presented~\cite{Gamberg:2010tj}: this leads to a process dependence of the Sivers function and to a an almost opposite SSA value in $pp\to\pi X$ compared to that obtained in the GPM.

Another fruitful application of the GPM is the study of azimuthal distribution of pions inside a jet in $pp$ collisions. Compared with inclusive pion production, the Sivers and Collins contributions can be disentangled by properly projecting out, in close analogy to SIDIS, the various azimuthal modulations~\cite{D'Alesio:2010am}. Furthermore, it has been shown that this process could allow a clear test of the process dependence of the Sivers function~\cite{D'Alesio:2011mc}.

TMDs have opened a new window on our comprehension of the nucleon structure.
They require challenging extensions of the standard approach used for collinear PDFs, leading us to a deeper understanding of QCD. Universality and factorization exhibit novel and more puzzling features and only the close interplay between lepton-nucleon scattering and hadronic collisions will allow us to fully explore these issues.

\acknowledgments
The author thanks the organizers of this very interesting and fruitful workshop for their kind invitation, and F. Murgia for carefully reading this contribution.
He also acknowledges partial support by Italian MIUR under PRIN 2008, and by the European Community under the FP7 grant agreement No.~227431.

\end{document}